# Different ways of dealing with Compton scattering and positron annihilation experimental data.

G. Kontrym-Sznajd and M. Samsel-Czekała

*W. Trzebiatowski Institute of Low Temperature and Structure Research, Polish Academy of Sciences, P.O. Box 1410, 50-950 Wrocław 2, Poland*

gsznajd@int.pan.wroc.pl and samsel@int.pan.wroc.pl

Different ways of dealing with one-dimensional (1D) spectra, measured e.g. in the Compton scattering or angular correlation of positron annihilation radiation (ACAR) experiments are presented. On the example of divalent hexagonal close packed metals it is shown what kind of information on the electronic structure one can get from 1D profiles, interpreted in terms of either 2D or 3D momentum densities.

2D and 3D densities are reconstructed from merely two and seven 1D profiles, respectively. Applied reconstruction techniques are particular solutions of the Radon transform in terms of orthogonal Gegenabauer polynomials. We propose their modification connected with so-called two-step reconstruction.

The analysis is performed both in the extended $p$ and reduced $k$ zone schemes. It is demonstrated that if positron wave function or many-body effects are strongly momentum dependent, analysis of 2D densities folded into $k$ space may lead to wrong conclusions concerning the Fermi surface. In the case of 2D ACAR data in Mg we found very strong many-body effects.

# 1 Introduction

One-dimensional (1D) spectra, measured either in the Compton scattering [1] or 1D angular correlation of positron annihilation radiation (ACAR) experiment [2], represent a double (plane) integral of 3D electron (or $e$-$p$) momentum density $\rho(p)$ in the extended space $p$. Generally, one can reproduce $\rho(p)$ from its plane projections – according to our knowledge there are six different reconstruction algorithms [3-8]. However, when $\rho(p)$ is a strongly anisotropic and spectra are measured with a high resolution, its reconstruction demands a large number of profiles (measured for very particular directions $p_z$ [9]). Thus, for 1D data, we recommend reconstruction of 2D density, shortly described in the next Chapter (for more details see [10,11]), proposing also some modification of reconstruction techniques under applying to CPs.

On the example of divalent hcp metals we demonstrate what kind of information on the electronic structure one can get from 1D profiles (both for Compton and positron annihilation data) interpreted in terms of either 3D or 2D densities in the extended $p$ and reduced $k$ zone schemes.

The free-electron Fermi surface (FS) of divalent hcp metals with the axial ratio $r = c/a$ lower than $r_c = (3.375\sqrt{3})/\pi = 1.860735$ consists of the following elements [12]: the 1st zone holes around H points (caps); 2nd zone hole monster; 3rd zone electrons around: $\Gamma$ (lens), L (butterflies), K (needles, called for Be cigars), and 4th zone electron pockets around L (see Fig. 1). In the case of Mg and Be the free-electron FS contains all these elements. For Cd the needles do not exist due to $r$ higher than the critical value $r_c$ at which they disappear. The free-electron FS for $r = 1.5633$ (as in Be) on the main symmetry planes is presented in Fig. 1. The "real" FS in Mg have the same pieces differing only in their dimensions [13-14], while the "real" FS of Be [11,15] contains holes neither in the 1st nor 2nd bands around H and the 2nd zone monster (called coronet) is very narrow at $\Sigma$ and reduced around T. There are also neither electron lens around $\Gamma$ in the 3rd band nor electron butterflies and electron pockets around L in the 3rd and 4th bands, respectively. The electron needles in the 3rd zone are much larger then for the free-electron model. The real FS's in $d$-electron metals Zn and Cd (the last presented in Fig. 1) also differ (but less in comparison with Be) from the free-electron FS's [13,16,17]. They do not contain electrons around L and the 2nd zone monster is much reduced (in Cd it forms six separate shits around K). In Zn the existence of needles around K is still controversial [17] - it could depend on changes of lattice constants with e.g. temperature or pressure. However, even though they should exist, they would be very small, beyond the momentum resolution of both Compton or ACAR experiments.

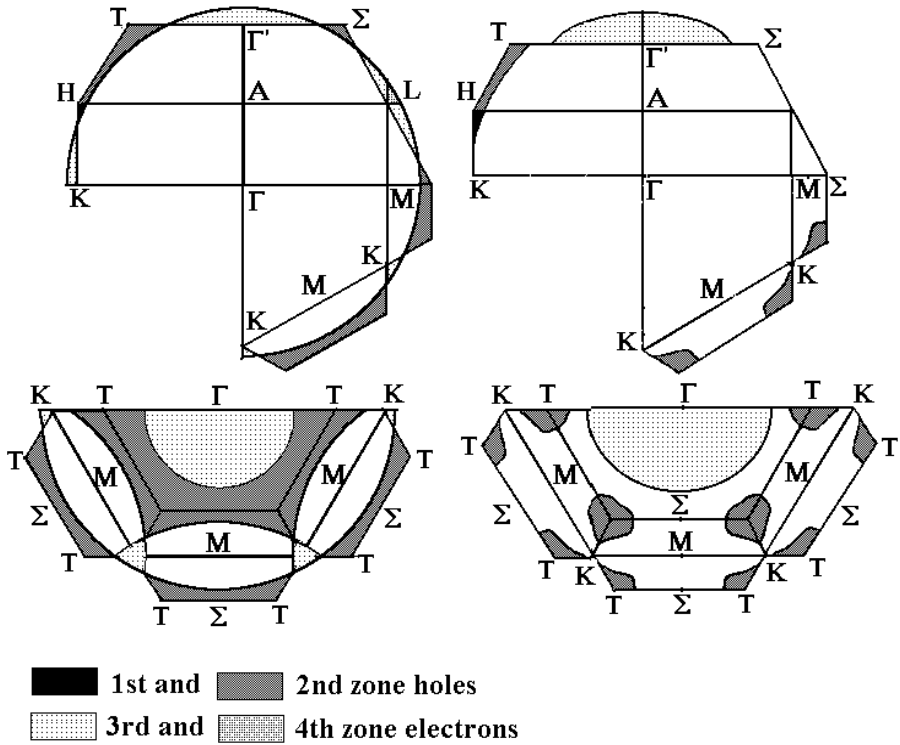

**Fig. 1** The free-electron FS of Be (left panel) and "real" FS of Cd (right panel). The FS's are presented in ΓAK, ΓAM and ΓMK planes in the extended zone (upper part) and in the ΓMK plane in the reduced zone (lower part) where some elements are additionally repeated.

## 2 Applied Techniques

1D spectrum, being a plane integral of 3D density $\rho(\boldsymbol{p})$, can be treated also as a line integral of the 2D momentum density $\rho^L$

$$J(p_z) = \int_{-\infty}^{\infty}\int_{-\infty}^{\infty} \rho(\boldsymbol{p}) dp_x dp_y = \int_{-\infty}^{\infty} \rho^L(p_z, p_y) dp_y , \qquad (1)$$

defined as

$$\rho^L(p_z, p_y) \equiv J(p_z, p_y) = \int_{-\infty}^{\infty} \rho(\boldsymbol{p}) dp_x. \qquad (2)$$

Generally, reconstruction of $\rho(\boldsymbol{p})$ from 1D profiles demands a large number of spectra, meanwhile $\rho^L(\boldsymbol{p})$ can be reconstructed much more easily and with higher precision. Such a conversion 1D $\Rightarrow$ 2D is very reasonable because there are no doubts that 2D density contains more details than 1D one. Such an approach has already been successfully applied to Compton profiles (CP's) in chromium [18] and beryllium [11].

To reproduce $\rho^L$ efficiently from the smallest number of projections, one should measure $J(p_z)$ for $p_z$ perpendicular to the [001] direction (to the main rotation axis). It allows to reconstruct $\rho^L$ with $\boldsymbol{L}$ along [001], from the very small number of profiles (we showed that in yttrium three profiles were quite sufficient to reproduce properly $\rho^{L=[001]} \equiv \rho^{001}$ [10]). Here different computerized tomography techniques [19,20] can be applied (conversion 1D $\Rightarrow$ 2D is the same mathematical question as 2D $\Rightarrow$ 3D). We used the Cormack method [21]. Having two profiles $J(p_z)$, with $p_z$ along $\Gamma K$ and $\Gamma M$ (directions perpendicular to the hexagonal axis of the 6$^{th}$ order), we can get two density components $\rho_0^{001}$, $\rho_6^{001}$ and $\rho^{001}$ from the following equation:

$$\rho^{001}(p, \varphi) = \rho_0^{001}(p) + \rho_6^{001}(p) \cos(6\varphi) . \qquad (3)$$

The same procedure was employed to reconstruct 3D densities from 2D ACAR spectra [22] while in order to reproduce 3D densities from 1D profiles, we applied the Jacobi polynomials algorithm [8]. All these techniques are equivalent being particular solutions of the Radon transform in terms of orthogonal Gegenabauer polynomials [20,23].

However, there can be some question connected with a description of $J_l(p)$ by a finite polynomials series. Namely, each spectrum $J(p_z)$ must be described by one series in the whole momentum region (in the case of Compton profiles up to e.g. $p = 10$ a.u.). In the case of high-resolution spectra description of their details requires a very high number of polynomials to be used. However, because it is not profitable due to the statistical noise, we proposed a two-step reconstruction, shortly described below. The procedure can be applied only to an isotropic part of spectra, $J_0(p)$, allowing not only to reproduce details of the isotropic density $\rho_0(p)$ but also to eliminate significantly statistical noise propagated in the reconstructed densities – see Fig 2.

The idea of this two-step reconstruction (or $n$-step if the shape of $J_0(p)$ would be more complicated) is the following. $J_0(p)$ can be subdivided into two (or more) curves $J_0(p) = J_0^1(p) + J_0^2(p)$. $J_0^1(p) = J_0(p)$ for $p \geq p'$ and for $p < p'$ can be chosen freely but has to be much lower than $J_0(p)$ ($J_0^1(p)$ in contrast to $J_0(p)$ should represent the integrals of some density which is smooth and therefore can be described by a small number of polynomials. E.g. $J_0^1(p) = J_0(p')$ for $p' = p_F + \Delta p$ where $\Delta p$ is at least of the order of the experimental resolution. $J_0^1(p)$ for $p < p'$ (the same for $J_0^2(p)$) do not have any physical interpretation – they are only auxiliary curves in reconstruction of $\rho_0(p)$. This is a reason that such a procedure can be applied only for $J_0(p)$ - functions $J_{n\neq 0}(p)$ are particular (their minimum number of zeros is equal to $n/2$ [21]). However, because anisotropy of densities occurs for lower momenta, anisotropic components of spectra are considered always for shorter momenta where they can be described in detail by the one polynomials series.

Having functions $J_0^1(p)$ and $J_0^2(p)$, the method resolves itself to the following steps:

1. in the unit system $p_{max}=1$, $\rho_0^1(p)$ is evaluated for $J_0^1(p)$.
2. in the unit system $p'=1$, $\rho_0^2(p)$ is evaluated for $J_0^2(p)$.
3. coming back to the 1$^{st}$ unit system one gets:

    $\rho_0(p) = \rho_0^1(p)+(1/p')^k\rho_0^2((1/p')p)$ where $k =1$ for N=2 (conversion 2D$\Rightarrow$ 3D)

    and $k =2$ for N=3 (conversion 1D$\Rightarrow$ 3D).

For some model density $\rho_0(p)$, displayed in Fig. 2, we calculated its line projection $J_0(p)$, simulating statistical noise using a Gaussian random number generator with the standard deviation $\sigma = \sqrt{N}$ where $N$ denotes the total number of counts per sampling point (here for "experimental" statistics where $N$ at peak $J_0(p=0) = 50\ 000$). Next $\rho_0(p)$ was reconstructed by the standard [21] and modified Cormack's method.

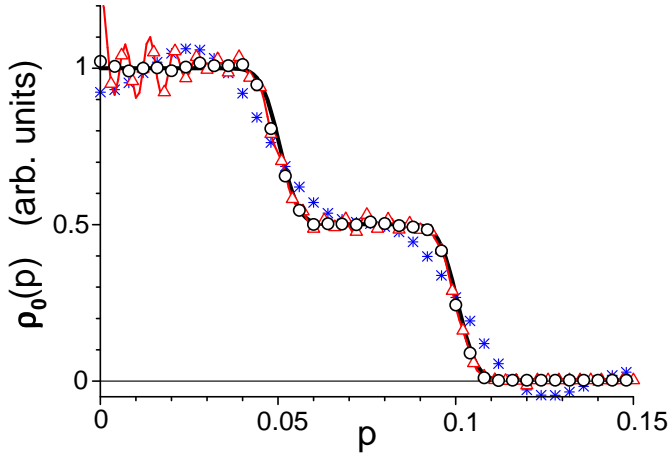

**Fig. 2** $\rho_0(p)$ reconstructed via one-step using 60 Chebyshev polynomials (stars) and two-step reconstruction, using 60 (line with open triangles) and 30 (open circles) Chebyshev polynomials for model spectra with noise, compared with the model $\rho_0(p)$ (thick solid line).

It is seen that 60 Chebyshev polynomials used in the one-step reconstruction are not sufficient to reproduce the model. It could be done by using much higher number of polynomials but in such a case there are strong fluctuations connected with the statistical noise (accumulated mainly for small momenta) – similarly as presented in Figure 2 for two-step reconstruction when 60 Chebyshev polynomials were used. However, by taking a smaller number (here 30 polynomials) in two-step reconstruction simulated error is almost completely smoothed (it demonstrates mean-squares approximation properties of orthogonal polynomials) and our model is reproduced perfectly.

Finally, to obtain a momentum density in the reduced space $k$, the Lock-Crisp-West (LCW) transformation was performed [24], i.e. a conversion from the extended $p$ to reduced $k$ space: $\rho(k) = \Sigma_G \rho(p=k+G)$ where $G$ are the reciprocal lattice vectors. If the influence of the positron wave function and many-body effects are ignored, it depends only on the electron occupation numbers ($n_j(k)$ = 0 or 1) in the $j^{th}$ band. Then the total contribution of all bands is equal to $\rho^\varepsilon(k) = \Sigma_j n_j(k) = n(k)$ where $n(k)$ denotes the number of occupied bands at the point $k$. In the case of the e-p densities $\rho_j(k) = n_j(k) f_j(k)$ where the function $f_j(k)$ depends on the electron state $|kj>$, even though correlation effects were neglected [25]. Particularly, small values of $f$ are expected for localized $d$-like states because an increasing localization of an electron state decreases the probability of e-p annihilation. So, if the character of various states $|kj>$ is strongly varying, their relative contributions to $\rho_j(k)$ may be essentially different. Nevertheless, as follows from theoretical calculations [25], the values of $f_j(k)$ are usually enough high to reproduce an observable jump of $\rho_j(k)$ if this quantity passes from one to another band.

this quantity passes from one to another band.

In the case of 2D electron density, $\rho^L(k)$ represents the line integral of $\rho(k)$ along direction $L$, i.e. it can be identified with the sum of line dimensions of the electron FS along $L$ (modified by $f_j(k)$). On the example of Mg we will show that one should be very careful studying the FS from $\rho^L(k)$ in such a case that function $f_j(k)$ is strongly momentum dependent.

Although, here we do not perform quantitative analysis, we estimated the influence of statistical noise on reconstructed densities. 2D ACAR spectrum for free electrons (with Kahana-like enhancement), simulating isotropic component of 2D ACAR data in Mg, was calculated for the same mesh of momenta (64×64) as in the experiment and convoluted with the experimental FWHM=0.1 a.u. For the spectrum we performed 20 different simulations of statistical noise, employing a gaussian random number generator with the standard deviation $\sigma = \sqrt{N}$ corresponding to the experimental statistics – in the case of 2D ACAR data for Mg [26], with 300 000 counts at peak. Then, for 2D spectra with noise we determined 20 different 1D model profiles and reconstructed 20 sets of 2D densities $\rho_0^{001}(p)$. Finally, the error distribution in terms of standard deviation $\sigma = \sigma[\rho_0^{001}(p)]$ was evaluated using estimators $\sigma[\rho(p)] = \sqrt{(1/M)\sum_{i=1}^{M}[\rho_i(p)-\bar{\rho}(p)]^2}$ where the average value of the densities with noise, $\bar{\rho}(p)$, is equal to the model density $\rho(p)$.

As a result we obtained the statistical error $\sigma[\rho_0^{001}(p)]$ is less than 0.6 % of $\rho_0^{001}(p=0)$ in the range of $p$ up to 0.1 a.u. and not higher than 0.2 % for the remaining momenta. Thus, in our analyses, the influence of the statistical noise on reconstructed 2D densities is to be neglected, the more so as we restrict our analysis to study only qualitative effects.

Here we would like to point out that such great statistics – 75 000 counts at peak for $J(p_z,p_y)$ (measured independently in 4 quarters $(p_z,p_y)$) corresponds to 3 770 000 counts at peak for 1D spectrum. Moreover, for ACAR data, where the core contribution is very small, such high statistics is almost totally for valence electrons.

## 3 Results

Maximum information on the electronic structure one can get having a sufficient number of spectra to reconstruct 3D densities. We show that for Cd 3D densities reconstructed either from two 2D ACAR or seven 1D ACAR profiles $J(p_z)$ (but for very particular directions $p_z$ as marked in Fig. 3) allow to obtain similar details of the FS.

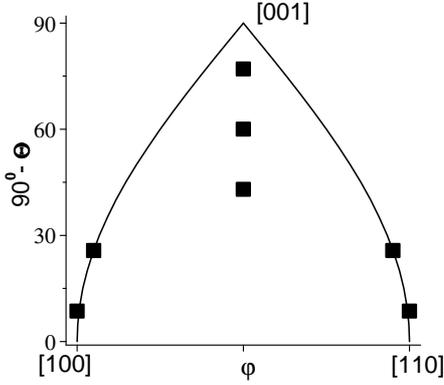

**Fig. 3.** The choice of special 7 directions $p_z(\Theta,\varphi)$ for the hcp lattice described by the spherical angles $(\Theta,\varphi)$ where five angles $\Theta$ are zeros of $P_{10}(\cos \Theta)$ (more details in [9]).

Densities $\rho(k)$ reconstructed from merely two 2D ACAR spectra [26], presented in Fig. 4, quite well reproduce the real FS of Cd, showing first of all a lack of electrons around L in the 3rd and 4th bands. If they existed, the densities around L would be higher than around Γ (electron lens in the 3rd band). The next feature is an existence of holes around H clearly visible in the 2nd band. Of course, if they exist in the 2nd band, they must exist also in the 1st band since AH line is on the border of these two bands (see Fig. 1). If these holes were larger, $\rho(k)$ should be lower than in the point K (electrons in the 1st band). However, because their sizes are smaller than experimental FWHM=0.1 a.u. as well as densities were reconstructed only from two 2D ACAR spectra, we are not able to study such subtle details.

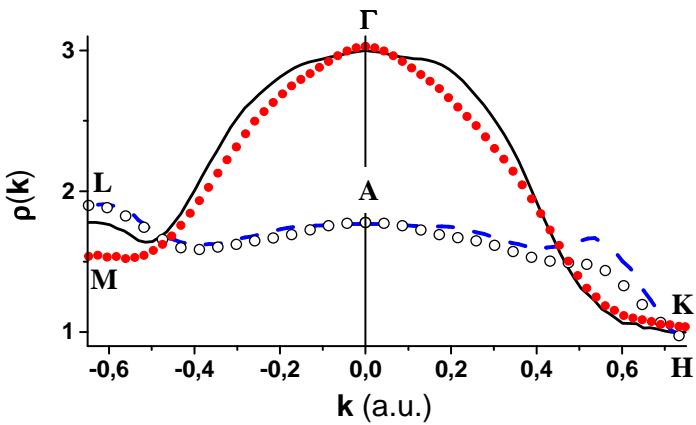

**Fig. 4.** Densities $\rho(k)$ in Cd, along the main symmetry lines MΓK (solid lines and circles) and LAH (dashed lines and open circles), reconstructed from two 2D ACAR (lines) and seven 1D ACAR (symbols) spectra, the last for directions displayed in Fig. 3.

Reconstruction of $\rho(\bm{p})$ from two 2D ACAR spectra is equivalent to the assumption that the anisotropy of 3D density $\rho(\bm{p})$ can be described by the lattice harmonics $P_l^{|m|}(\cos\Theta)\cos(m\varphi)$ with $m = 0$ and 6. When performed independently on each plane perpendicular to the hexagonal axis, there is no limitation as concerns $l$, i.e. $\rho_0(p,\Theta)$ and $\rho_6(p,\Theta)$ are described by the lattice harmonics $P_l(\cos\Theta)$ and $P_l^6(\cos\Theta)\cos(6\varphi)$, respectively, with arbitrarily high $l$. However, if there are no harmonics with $m=12$, one can expect that reconstructed densities from 1D profiles will give similar results (of course, always more smeared due to more limited number of lattice harmonics) for nine plane projections - if the tenth harmonics $P_{12}(\cos\Theta)$ was substantial, the next harmonics of the same order $l=12$ but with $m=6$ and 12 could be essential, too. Our tests showed that densities reconstructed even from 7 projections 1D ACAR reproduce densities reconstructed from two 2D ACAR spectra (see Fig. 4). Results for 7 are practically the same as for 9 plane projections, while for 5 are more smeared though they still reproduce main features of the FS in Cd.

Next, we present results for $\rho^{001}$ reconstructed from two 1D profiles in divalent hcp metals. 1D ACAR spectra for Mg and Cd were created from 2D ACAR data [26] measured for $p_z$ along directions [100] and [110] and with an overall momentum resolution (FWHM) of 0.1 a.u. From seven and two 1D profiles 3D and 2D $e$-$p$ momentum densities, i.e. $\rho(\bm{p})$ and $\rho^{001}(\bm{p})$, were reconstructed. Next, corresponding LCW densities $\rho(\bm{k})$ and $\rho^{001}(\bm{k})$ were created and compared with 2D electron densities, $\rho^{001}(\bm{k})$, reconstructed also from two Compton profiles in Be [11].

The isotropic component of the densities, $\rho_0^{001}(p)$, reconstructed from the average of two 1D ACAR spectra, $J_0(p)$, is presented in Fig. 5. It shows that in the case of positron annihilation spectra the contribution of core states in Mg is very small while it is relatively high in Cd. This is connected with the fact that in Cd there are $4d^{10}$ electrons, which (according to band structure results [13]) are above the bottom of the conduction band minimum. So they are "seen" by a positron with a higher probability than typical core states.

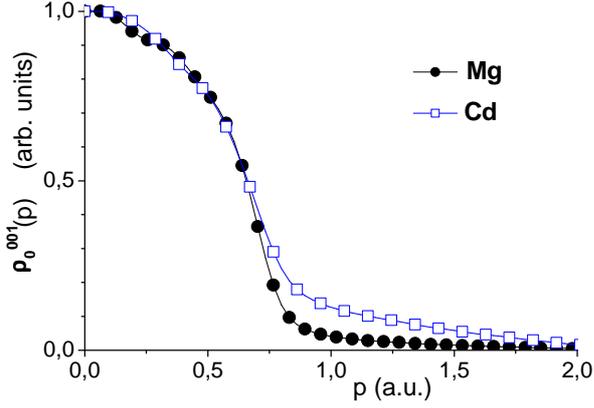

**Fig. 5**

The isotropic component of 2D densities, $\rho_0^{001}(p)$, [normalised to $\rho_0^{001}(0) = 1$] reconstructed from two 1D ACAR spectra in Mg and Cd.

The anisotropy of the reconstructed 2D densities (equal to $\rho_{\Gamma K}^{001}(p) - \rho_{\Gamma M}^{001}(p)$ and represented by $2\rho_6^{001}$ in Eq. (3)) is displayed in Fig. 6. For momenta less than 1 a.u. it contains information on the anisotropy of the FS: the maximum around KH line in Mg is connected with the reduction of the 1st zone holes around H at the cost of the reduction of the 3rd and 4th zone electrons around L. In Cd the anisotropy is similar, although its maximum occurs along the ML line, which could reflect the lack of electrons around L points (butterflies and pockets), clearly observed in *k* space for both 3D (Fig. 4) and 2D (Fig. 7) densities. Anisotropy above the Fermi momenta is connected with umklapp components.

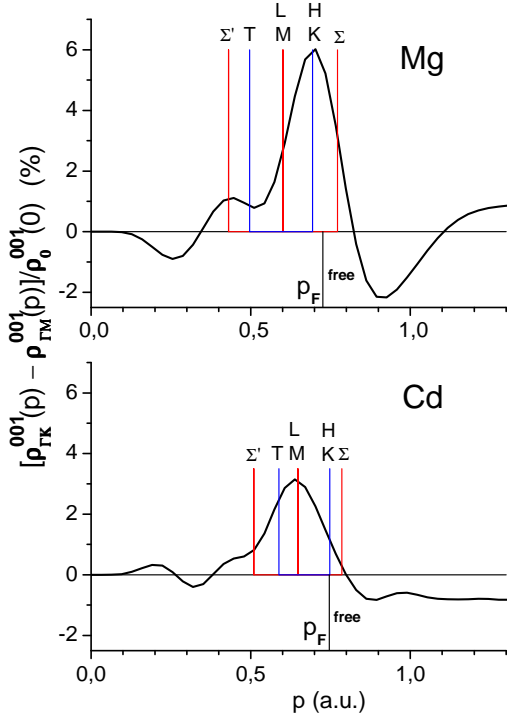

**Fig. 6**

Differences between $\rho^{001}(p)$ [in % of $\rho_0^{001}(p=0)$] for momenta along ΓK and ΓM, reconstructed from two 1D ACAR spectra, for Mg and Cd.

It is obvious that it is impossible to obtain the shape of the FS from the density $\rho(p)$ itself (the same for $\rho^J(p)$), due to the fact that $\rho(p)$ is not constant on the FS and represents a sum of contributions from all occupied bands, not only those crossing the Fermi energy. Thus, in order to map the FS, usually the LCW folding [24] is performed. Densities $\rho^{001}(k)$ for divalent hcp metals for the free-electron model and experiments are presented in the left and right part of Fig. 7, respectively. "Experimental" 2D densities $\rho^{001}(k)$ in Mg and Cd were obtained from $\rho^{001}(p)$ reconstructed from two 1D ACAR while for Be from two theoretical (convoluted and Lam-Platzman corrected [27]) and two experimental Compton profiles.

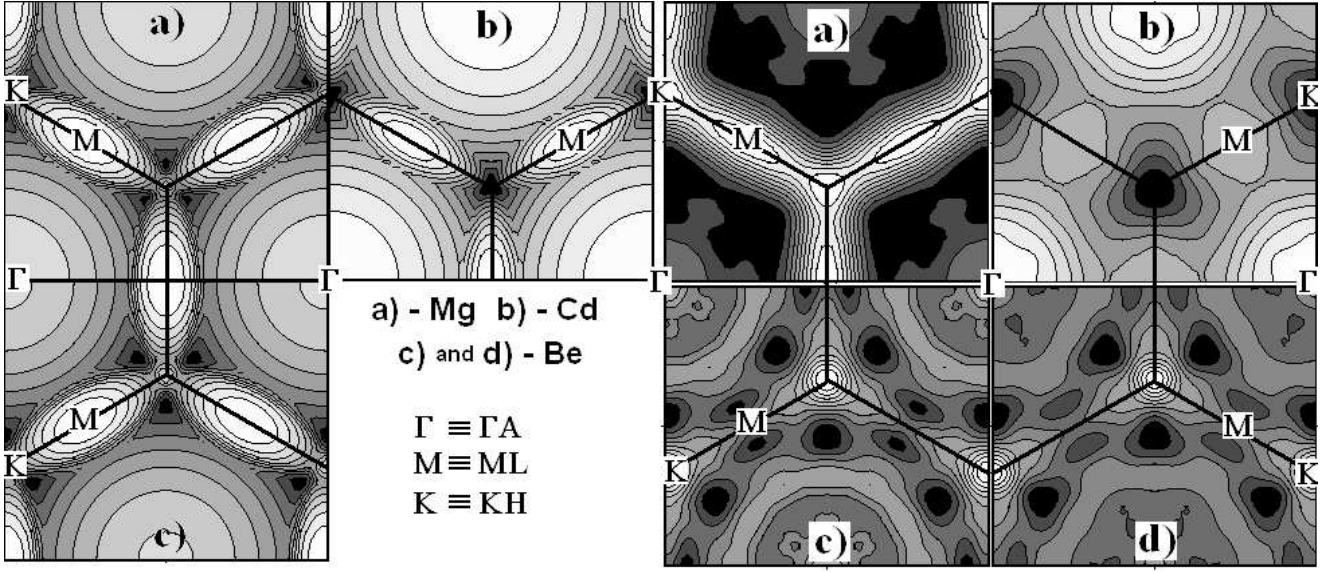

**Fig. 7.** Free electron 2D densities $\rho^{001}(k)$ in divalent hcp metals (left side) compared with real densities (right side), reconstructed from two 1D spectra: for Mg (a) and Cd (b) from two 1D ACAR profiles and for Be (c,d) from two experimental (c) and convoluted theoretical (d) CP's. Data for Be are the same as in Fig. 5II(b,c) [11] and all results are drawn separately (in different scales of grey shades) by 11 contourlines.

It is seen that free-electron densities $\rho^{001}(k)$ in all divalent hcp metals are very similar, which is not a case for the "real" densities. For Be interpretation is essentially simplified due to corresponding results obtained for theoretical Compton profiles [28]. Almost perfect agreement between theory and experiment confirms all details of the theoretical band structure calculations [28]. In the case of Cd we observe a lack of butterflies and pockets when compared to the free-electron densities. More surprising are results obtained for Mg, first of all for the 3rd band electron lens around Γ. Namely, it seems that real $\rho^{001}(k)$ reconstructed from two 1D ACAR spectra (Fig. 7a – right side) do not contain the lens or it is much reduced in comparison with the free-electron model (Figs. 7a – left side). However, it is only seeming – the lens is clearly seen in $\rho^{001}(\boldsymbol{p})$ ($\rho_0^{001}(p)$, displayed in Fig. 5, is not flat for small momenta as for Be [11] where there is no lens around Γ) as well as in 3D densities, both in $\boldsymbol{p}$ and $\boldsymbol{k}$ space (e.g. Figs. 4 and 6 in Ref. [22]). Our tests, which results are displayed in Figs. 8 and 9, point out that this behaviour can be explained by both strongly momentum dependent e-p correlations as well as smearing (resolution and e-e correlation) effects. In order to simulate them, we added the Kahanalike enhancement $[\varepsilon(p) = 1 + 0.15(p/p_F)^2 + 0.25(p/p_F)^4$ for $p \leq p_F = 0.7264$ a.u.] to the free-electron model densities, smeared next by the experimental resolution FWHM = 0.1 a.u.

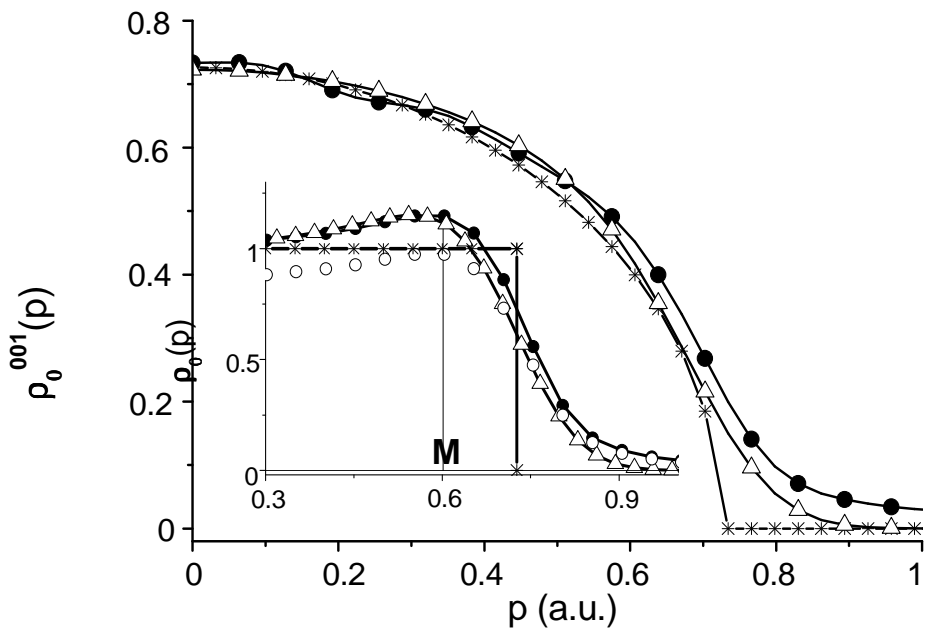

**Fig. 8** Isotropic components $\rho_0(p)$ and $\rho_0^{001}(p)$ in Mg, reconstructed from 1D ACAR spectra (solid lines with solid circles), compared with models: free electrons (solid lines with stars) and free electrons with Kahanalike enhancement, convoluted by FWHM=0.2 a.u. (solid lines with open triangles). Open circles show renormalized experimental densities.

The corresponding 2D densities were much closer to the experimental $\rho^{001}(\boldsymbol{k})$ but there was still large disagreement between model and experiment. So, we took an effective FWHM=0.2, simulating both some part of e-e correlations and experimental resolution. Now the corresponding $\rho_0(p)$ and experimental $\rho_0(p)$ have comparable smearing at the FS but these densities are shifted from each other. However, $\rho_0^{001}(k)$ are almost identical (see results presented in Fig. 9) showing that the "seeming" reduction of the lens around Γ is connected with strongly momentum dependent *e-p* correlations and smearing effects.

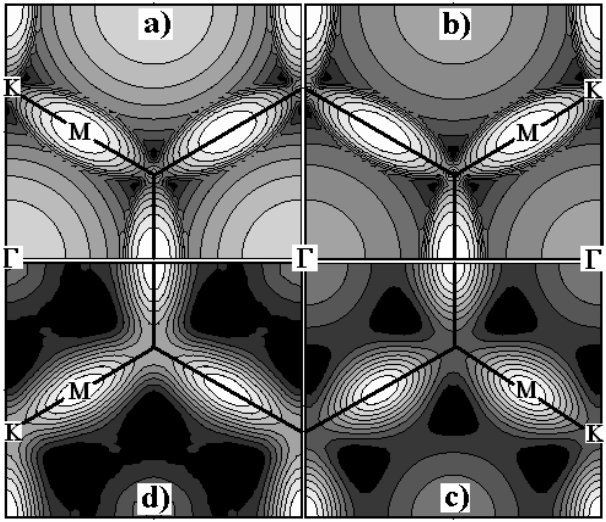

**Fig. 9** $\rho_0^{001}(k)$ in Mg:

(a) – free electrons; (b) – free electrons with Kahanalike enhancement; (c) – the same as in (b) but after convoluting with FWHM=0.2 a.u.; (d) - reconstructed from two 1D ACAR spectra.

Here we would like to notice the following. In order to compare the momentum dependence of theoretical and experimental densities we normalized them to the same value at *p*=0. It is connected with the fact that in the ACAR experiment we do not measure absolute values of the momentum density *ρ(**p**)*. Due to the presence of the positron their normalization is somewhat ambiguous (contrary to the Compton spectra which areas are equal to the number of electrons). Of course, normalisation does not influence the LCW results which show only relative differences between $\rho(k)$ and $\rho(k')$.

We checked that obtained parallel shift between experimental and smeared model $\rho_0(p)$ densities (line with triangles and solid circles in figure 8) is not connected with wrong elaboration of experimental data. So, because the FS of Mg must contain 2 electrons (i.e. the average Fermi momentum should be the same as for free electrons), we eliminated this shift by renormalizing experimental $\rho_0(p)$, multiplying it by the factor= 0.85. Results (open circles in Fig. 8) suggest that experimental densities contain strong *e-e* correlations, not seen before, when all data were normalized to the same values at *p*=0. The effect of *e-e* correlations in Mg, much stronger than described by the Lam-Platzman correction, was observed in the Compton scattering profiles [29].

An influence of a positron on 2D densities in $k$ space was observed for Cr [18]. Dugdale et al. [18] got discrepancy between electron ($\rho^{011}(k)$ reconstructed from seven 1D Compton profiles) and *e-p* densities (2D ACAR data folded into the 1$^{st}$ Brillouin zone) for Cr. However, because in Cr (contrary to Mg) the effective enhancement factor is not momentum dependent [30], it seems that the effect observed in Cr is connected with either positron wave function or selectivity of enhancement - in both cases selectivity of positron annihilation with *d*- or *s*-like states.

There are interesting results obtained lately for UGa$_3$ [31]. From 3D densities reconstructed from 2D ACAR data, authors estimated the FS getting direct evidence of 5*f*-electrons itinerancy in UGa$_3$. They also showed that, due to a strong positron wave function effects in this uranium compound, it would be difficult to estimate the FS even from 3D *e-p* densities $\rho(k)$ without the knowledge of corresponding theoretical *e-p* densities.

# 4  Conclusions

In the Compton scattering experiment one probes electron densities not disturbed by a positron, which constitutes its advantage in electronic structure studies. However, in this experiment one measures:

a) densities from both core and valence electrons (in the case of ACAR data the core contribution, reduced by a positron, is very small, *i.e.* the total experimental statistics is mostly connected with valence states).

b) plane (not line) integrals where information on $\rho(p)$ is more entangled.

Because reconstruction of densities from plane projections requires a lot of profiles, in the case of Compton scattering spectra their analysis was performed mainly for $J(p_z)$ (e.g. [28,32]). However, development of synchrotron sources and the possibility of measuring many high-resolution Compton profiles makes this technique promising for the future [33,34], the more so as there is even pos-sibility of measuring directly 3D electron densities using (γ,eγ) technique [35] (up to now limited mainly due to statistics).

We showed that for Cd seven 1D profiles (but measured for specific directions) allow to reconstruct essential features of 3D densities and draw the FS. We also advise simple conversion from 1D to 2D densities because it can be done for a small number of profiles, yielding information on the FS not so strongly entangled as in 1D spectra. Moreover, such results can be compared with corresponding 2D ACAR spectra, which allows to examine with greater care both electronic structure and many-body effects (such analysis was even performed for measured directly 3D densities [35]). Nevertheless, as shown in this paper, the analysis of 2D spectra in the reduced zone ($\rho^L(\mathbf{k})$) may lead to wrong conclusions concerning the FS if many-body or positron wave function effects are strongly momentum dependent. So, such an analysis could be done very carefully, first of all in the case of ACAR data.

A disadvantage of the positron annihilation experiment in comparison with the Compton scattering technique sometimes becomes its advantage - when it allows to qualify a degree of the electron localization. E.g. 2D ACAR spectra for Cd show that $4d$ electron bands are not strongly localized as typical core states (they are above the bottom of the conduction band minimum).

We propose also two-step reconstruction of isotropic components of densities for both line and plane projections. Such a procedure (for techniques employing orthogonal polynomials) is particularly reasonable for Compton spectra where densities are measured in a long momentum range.

**Acknowledgements** We are very grateful to the Polish State Committee for Scientific Research (Grant No. 2 P03B 012 25) for financial support.